\documentclass[10pt, conference]{IEEEtran}
\ifCLASSINFOpdf
  \usepackage[pdftex]{graphicx}
\else
\fi

\usepackage{balance}

\begin{document}
%
\title{Field Testing of Software Applications}


\author{\IEEEauthorblockN{Luca Gazzola}
\IEEEauthorblockA{Universita' degli Studi di Milano-Bicocca
Milano, Italy\\
luca.gazzola@disco.unimib.it\\
Supervisor: Leonardo Mariani}
}


%


\maketitle

\begin{abstract}
When interacting with their software systems, users may have to deal with problems like crashes, failures, and program instability. Faulty software running in the field is not only the consequence of ineffective in-house verification and validation techniques, but it is also due to the complexity and diversity of the interactions between an application and its environment. Many of these interactions can be hardly predicted at testing time, and even when they could be predicted, often there are so many cases to be tested that they cannot be all feasibly addressed before the software is released.  

This Ph.D. thesis investigates the idea of addressing the faults that cannot be effectively addressed in house directly in the field, exploiting the field itself as testbed for running the test cases. An enormous number of diverse environments would then be available for testing, giving the possibility to run many test cases in many different situations, timely revealing the many failures that would be hard to detect otherwise. 

\end{abstract}


%
\IEEEpeerreviewmaketitle

\section{Testing Software in the Field}

Achieving high quality is mandatory in modern software applications, but, despite intensive in-house testing sessions, organizations still struggle with releasing dependable software. Faulty applications running in the field are the source of several problems, including higher maintenance costs, reduced customers satisfaction, and ultimately loss of reputation and profits.

Studying the reason why several faults are not detected with in-house testing is of crucial importance to mitigate their occurrence in the field. Although the faults present in the field could be the result of a poor in-house testing process, there are many faults that are objectively hard to detect in house, even using state of the art methodologies and techniques.  

This intuition has been confirmed by an analysis that we performed on multiple real software failures experienced in the field by the end-users. Our analysis shown that a significant proportion of the faults that cause field failures have specific characteristics that make them extremely hard to be detected in house. In particular, out of all the failures that we analysed, approximately only 30\% of them could be attributed to bad testing practices, while the remaining 70\% should be attributed to an interaction between the software under test and the field that is objectively difficult or even impossible to test in house. 
The impossibility to thoroughly test software systems in house has been already recognized in mature companies. For instance, Netflix engineers inject faults in the field to analyse the impact of faults that cannot be studied in house in a simulated environment~\cite{Basiri:Netflix:ISSRE:2016}. 

The vision of this Ph.D. thesis is that the end-user environment could be exploited as testbed for running the many test cases that cannot be executed in house, because of both the limited resources available for testing and the challenge of recreating in house the same environment that is available in the field. There are several important benefits related to the capability of exploiting the field as part of the testing process. If the application to be tested is popular enough to be installed in many devices and computers (e.g., hundreds or thousands), these many instances would offer unique opportunities in terms of range of situations and configurations that could be tested in parallel. Moreover, regardless the number of installations of a same application that are available, testing an application in the field gives the unique opportunity of testing the software when used in its real environment, with real data. Finally, test cases could be executed perpetually, not only to test an application that has been just installed, but also to test the software while the environment, the users, and the data evolve. 

Successfully deploying the capability of executing test cases while the software is running in the field, namely the capability of doing field testing, even enables the possibility to identify potential failures before they are experienced by the end-users.



\section{Key challenges in Field Testing}
\label{sec:challenges}


Field testing is inherently different from traditional in-house testing and poses a new set of challenges, the ones about the \emph{test strategy}, that is, when and how to test the software in the field, and the ones about the \emph{non-intrusiveness} of the testing process, that is, how to run the test cases without affecting the user data and user processes, with negligible impact on the user experience.

\subsection{Test strategy}

We identified four key elements that should be part of a well-defined test strategy.

\smallskip \emph{Test obligation}. The test obligation defines \emph{which} behaviors should be tested in the field. Since field testing is executed after in-house testing, field testing should focus on complementary aspects, that is, on those cases that have not been tested in house. Moreover, field testing should target those functionalities that cannot be effectively and efficiently tested in house, such as, functionalities that depend on field entities or functionalities whose execution space can combinatorially explode. An example of the former functionality is a routine that uses an external DBMS, which requires interacting with a specific driver and a specific database configuration. An example of the latter functionality is an application for writing documents, which may display some text incorrectly only when using specific characters of a specific size and of a specific font type.


\smallskip\emph{Test opportunity}. The test opportunity defines \emph{when} the software application running in the field should be tested. When testing is performed in house, the testing procedures are usually started in accordance with the development activities (e.g., a change in the source code). The activation of testing procedures for software running in the field should be based on different aspects. In particular we identified two key elements. The first one is the software state, which could be recognized as relevant for testing. For example, the test cases for a spreadsheet application might be executed when the user opens a spreadsheet with thousands of formulas and millions of data values, assuming that a real spreadsheet of this size is a relevant subject for the testing procedure. The second element is the available resources, which should be sufficient to run the test cases without impacting the user experience. For instance, if the system is already performing a cpu intensive computation, it might be a bad idea to activate an automatic testing procedure.

\smallskip\emph{Test generation strategy}. The test generation strategy defines \emph{how} to obtain the test cases that should be executed in the field. We foresee two different types of strategies: static and dynamic. Static test generation strategies consist of implementing in house the test suites specifically designed to cover potentially interesting situations in the field. In this case the test suite would be static, and the test cases that must be executed would be selected dynamically based on the state of the execution. Dynamic test generation strategies should instead generate or update test cases directly in the field. This can be done by developers, based on problems reported by end users, or automatically, for example by mutating the available test cases or even producing new tests from scratch.

\smallskip\emph{Test oracle}. The test oracle defines \emph{what} the expected behavior of the software under test is. This information is necessary to detect failures. There are at least two main approaches to obtain a test oracle that can detect failures, beyond program crashes. An approach consists of defining a generic oracle for the application, covering at least the functionalities that should be tested in the field. An alternative approach is associating oracles with test cases. The former approach is more generic and can serve many purposes, including the detection of  failures revealed by automatically generated test cases, but it is more expensive. The latter approach is cheaper to specify because it covers just the specific cases encoded in the test cases, but it is harder to generalize to executions different from the ones represented in the test cases, and thus it might be difficult to reuse for the automatically generated test cases. 

\subsection{Non-intrusiveness}

We identified two key elements that should be considered about non-intrusiveness.

\smallskip \emph{Isolation}. A software application running in the field interacts with the elements that are available within its environment (e.g., the resources). Isolating field testing requires the guarantee that the application under test does not produce any disruptive effect while tested, that is, it cannot cause any loss of data and cannot impact on the other processes running in the field. This is a fundamental property to make the field testing technology acceptable by end users.

\smallskip \emph{Overhead}. Running test cases in the field requires the consumption of additional resources, definitely cpu cycles and memory, but also I/O and access to the network. The field testing procedure must guarantee that any additional resource consumption is rarely and hardly noticeable by the users of the application.

\section{Research Methodology}

The research methodology adopted in this Ph.D. thesis consists of three main phases. The first phase, namely \emph{analysis of field failures}, concerns with the analysis of the failures reported from the field by users of software systems. 
The second phase, namely \emph{identification of test strategies}, concerns with the definition of a test strategy aimed to reveal the faults that can be hardly revealed in house, based on the outcome of the first phase. The third phase, namely \emph{definition of the test infrastructure}, concerns with the design and development of a prototype infrastructure that supports the execution of the test strategies defined in the second phase, guaranteeing the non-intrusiveness of the testing process.

The first phase of the research has been completed, and we are now facing the second and third phases that will be developed in parallel. In the rest of this section, we discuss each phase, and the results that have been obtained, when available. We conclude by sketching the validation plan.


\subsection{Analysis of Field Failures}

To understand the characteristics of field failures, we focused on two main research questions:  


\begin{enumerate}
\item What are the reasons why software faults are not detected at testing time?
\item What are the field elements typically involved in a field failure?
\end{enumerate}

To answer these questions we manually inspected 119 bug reports\footnote{We actually inspected more than 400 bug reports, but we have been able to extract useful information about the fault for only 119 bug reports.} from three different open source systems: three Eclipse plugins~\cite{Subversive,EGit,EclipseLink}, OpenOffice~\cite{OpenOffice} and Nuxeo~\cite{nuxeo}. We discovered that only $30\%$ of the faults can be attributed to weak testing, while $70\%$ of the faults would be hard to reveal in house for one of the following four reasons, that we defined to answer the first research question:

\smallskip \noindent \emph{Impossible to test (ItT)}: it is impossible to replicate in house the environment that produces the fault. 

\smallskip \noindent \emph{Lack of information about the application (LoIA)}: the application fails for an undocumented specific case, thus testers have too little information to design a test case that covers the fault in-house. 

\smallskip \noindent \emph{Lack of information about the environment (LoIE)}: the application fails for a specific configuration of the environment that in principle should not have any effect on the application. 

\smallskip \noindent \emph{Combinatorial explosion (CE)}: the application fails for a specific case out of a huge set of possible cases that cannot be feasibly tested in house. 
%
%

\begin{figure}
\centering
\includegraphics[width=7.5cm]{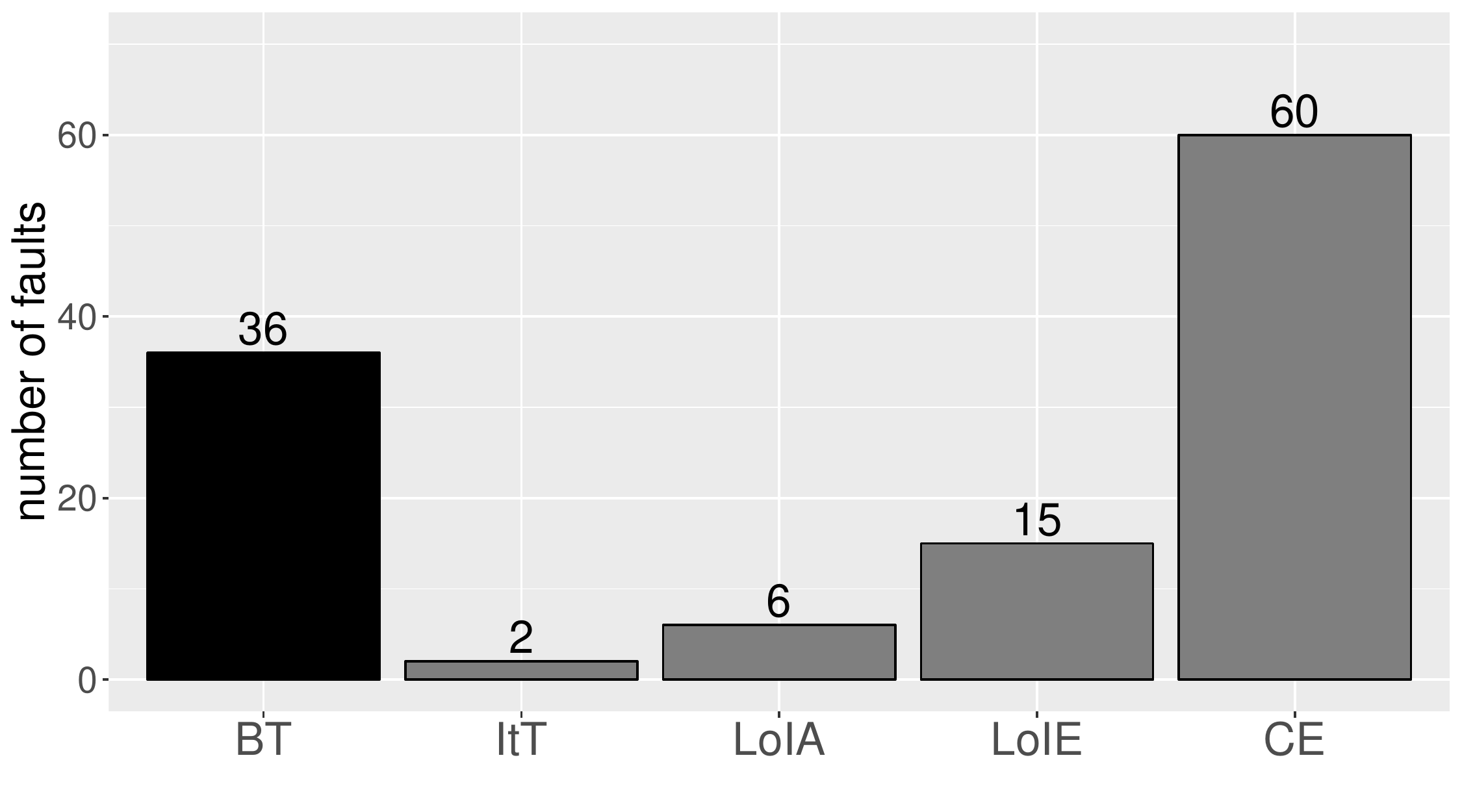}
\caption{RQ1: reasons why faults could not be revealed at testing time}
\label{fig:rq1}
\end{figure}
%

\smallskip Figure~\ref{fig:rq1} shows the relative distribution of these four causes for the bug reports that we analyzed, plus the bad testing (BT) cases. Faults related to combinatorial explosion is the major cause of field faults. We observed only two faults that were impossible to test in house. We expect this result to be influenced by the set of systems considered in our analysis. Probably the set of cases impossible to test in house would be higher if considering multi-user distributed systems.


To answer the second research question we checked how often a specific field element, potentially in a specific state, is necessary to observe a failure. We observed  that 79\% of field failures depend on field elements. The most frequent ones are resources (e.g., files), which appear in 54\% of the cases, and operating systems, which appear in 21\% of the cases.

These two results jointly show that (1) there are many faults that can be hardly revealed in house for reasons different than a weak testing process, and (2) the testing procedure must exploit elements present in the environment to successfully reveal these faults. 

%
%

\subsection{Identification of the Test Strategy}

In this phase, we exploit the knowledge gained with the initial study to distill a set of \emph{test strategies} for specific classes of field failures. Each test strategy consists of executing a predefined test activity when a given test opportunity is observed. According to our observations, many of the failures can be clustered into similar failure scenarios, which consist of observing specific operations when certain field elements are in a given state. Failure scenarios let us specify the failure opportunities that should be considered when performing field testing.

So far we identified several opportunities, an example is the \emph{modified resources location}. Many applications must access to resources available in the field when executed, such as files and databases. Sometime these resources are moved to a new location by an independent software or by the user. If the application running in the field is not robust against resource relocation, changing the location of resources might threat the correctness of the application. 

The test strategy for the modified resource location opportunity consists of generating test suites that cover several usage scenarios for all the resources used by an application. At runtime we periodically check if resources are moved, and, in case a relocation is detected, the test suite exercising the uses of that resource is automatically executed to check if the application can tolerate the new configuration.

%
%
%
%
%

We are currently working on more structured definition of test opportunities and corresponding test strategies, with the aim of covering a representative set of failures. 

\subsection{Definition of the Test Infrastructure}

In order to be able to deploy field testing, we need a suitable infrastructure providing facilities for maintaining the test suites available in the field, and collect the results produced by the execution of the test cases.

Since test strategies require the ability to detect when some test cases should be executed, the infrastructure must include lightweight monitoring capabilities for detecting the test opportunity, and should also be able to select or generate the relevant test cases.

A relevant part of the infrastructure should be devoted to guaranteeing the non-intrusiveness of the field testing process. To achieve isolation, we intend to exploit virtualization technologies and containers, such as Docker~\cite{docker}, to efficiently wrap a software application in a complete environment. 

A technical solution alternative to virtualization might be based on platforms for N-version executions, which consist of platforms that can run multiple instances of a program in parallel, keeping them synchronised through the use of monitors. Assuming to have enough resources to run multiple instances of a software application, this solution enables efficient field testing activities. An architecture supporting this form of N-version execution has been proposed by Hosek and Cadar~\cite{hosek2015varan}.

\subsection{Validation}

In order to evaluate this research, we will apply the defined field testing infrastructure to software applications from different domains: this will tell us if and how the application domain influences the success of field testing. Our measures of effectiveness will be based on the number of faults that can be discovered with the deployed infrastructure, and on the number of tests that are generated and executed in the field. In addition, to assess the non-intrusiveness, we will measure the performance overhead (and more in general the resource consumption caused by the infrastructure), and we will evaluate the impact of the overhead on the user experience by designing studies with human subjects.

\section{Related work}

The approaches related to this Ph.D. work can be organized into three groups: studies about the characteristics of software failures, approaches to run test cases in the field, and testing and analysis techniques that exploit field data. In the following, we briefly discuss these classes of approaches.


Existing \emph{studies of real software faults} draw conclusions on characteristics such as the distribution of faults types~\cite{Hamill-TrendsInFaults-TSE-2009,Fan-NuclearFailures-SF-2013}, the locality of faults~\cite{Hamill-TrendsInFaults-TSE-2009}, the distribution of faults across source files~\cite{Ostrand-FaultDistribution-ISSTA-2002}, the root cause, that is, the development phase in which a fault has been introduced, and human error that caused the fault~\cite{Leszak-ClassificationOfDefects-JSS-2002}, the relation between fault types, failure detection, and failure severity~\cite{Hamill-FaultTypesDetectionSeverity-SQJ-2014}, and the evolution of faults during bug fixing~\cite{Meulen-FaultsFailureBehaviour-ISSE-2004}. However, none of these studies has focused on the causes of field failures and the reasons why failures have not been discovered at testing time, as well as the common characteristics of field failures, which is the starting point of our research.

The problem of running test cases in the field has been preliminarily studied by other researchers. The approach to \emph{field testing} that is closest to the work developed in this thesis is in-vivo testing~\cite{murphy2009quality}. In-vivo testing consists of deploying unit test cases that are executed with a user-defined probability while the application is running to detect failures. However, in-vivo testing does not take into account the specific elements (e.g., resources in the environment) that should be exploited in a field testing strategy to detect the faults that are hard to reveal in house, and does not address the problem of the non-intrusiveness of the testing process, which is assumed to be guaranteed by construction by the tests. Another attempt to move the testing process in the field is the work by Memon et al.~\cite{memon2004skoll}. However, they focus on moving acceptance testing to the field, which is a different task compared to designing a technique for revealing the faults that are hard to reveal in house.

Finally, there are analysis techniques \emph{exploiting field data}. For instance, residual test coverage monitoring~\cite{pavlopoulou1999residual} collects coverage data from the field to complement the results of structural testing obtained in-house. Recently, Ohmann et Al.~\cite{ohmann2016optimizing} proposed a novel approach to limit the amount of instrumentation necessary to collect coverage data, increasing the applicability of residual test coverage in the field.

Statistical program debugging has been also deployed in the field~\cite{jiang2007context,chilimbi2009holmes}. In these approaches, software applications are lightly instrumented so that they can send execution profiles to a central database. Statistical debugging is then performed on the profiles gathered from the field to identify the likely locations of the faults that have been activated in the field. Although debugging is a different task than testing, the distributed infrastructure built to monitor and collect data from applications running in the field can be useful also to maintain test cases and gather results about the testing process.

\section{Conclusions}
Field testing can be a powerful solution for timely discovering the faults that are seldom detected with traditional in-house testing. Although this concept of testing is still in its infancy, the technical elements necessary to achieve it (e.g., testing techniques, virtualization environments, and distributed infrastructures) are mature enough to support this research. This Ph.D. thesis aims at exploring this novel concept of testing, introducing the concepts of test strategies and test opportunities, in addition to defining and developing an appropriate infrastructure for field testing.


\medskip \begin{small} \emph{Acknowledgments} This work has been partially supported by the H2020 Learn project, which has been funded under the ERC Consolidator Grant 2014 program (ERC Grant Agreement n. 646867).
\end{small}

\vfill



%

\bibliographystyle{abbrv}
\bibliography{bibliography}

\end{document}